\documentclass[twocolumn]{ijns}
\usepackage{graphicx,epsfig,color}
\usepackage[normal]{subfigure}
\usepackage{amsmath,amsthm,amssymb}
\usepackage{floatflt}
\usepackage{setspace}
\usepackage{rotating,algorithm,algorithmic}

\def\markboth#1#2{\def\leftmark{#1}\def\rightmark{#2}}

\newtheorem{definition}{Definition}
\newtheorem{theorem}{Theorem}
\newtheorem{lemma}{Lemma}

\newtheorem{corollary}{Corollary}

\title{\huge\bf Linear Complexity and Autocorrelation of two Classes of New Interleaved Sequences of Period $2N$ }
\author{Shidong Zhang$^{1,2}$, Tongjiang Yan$^{1,2}$\\
\medskip {\small\it(Corresponding author: Tongjiang Yan)}~\\
{\normalsize College of Science, China University of Petroleum, Qingdao, Shandong, China, 266580$^1$}\\
{\normalsize Key Laboratory of Network Security and Cryptology, Fujian Normal University, Fuzhou, Fujian, China, 350117 $^2$}\\
{\normalsize (Email:  yantoji@163.com)}\\
}

\date
\newpage
\begin{document}
\maketitle
\thispagestyle{headings}
\begin{abstract}
 The autocorrelation and the linear complexity of a key stream sequence in a stream cipher are important cryptographic properties. Many sequences with these good properties have interleaved structure, three classes of binary sequences of period $4N$ with optimal autocorrelation values have been constructed by Tang and Gong based on interleaving certain kinds of sequences of period $N$. In this paper, we use the interleaving technique to construct a binary sequence with the optimal autocorrelation of period $2N$, then we calculate its autocorrelation values and its distribution, and give a lower bound of linear complexity. Results show that these sequences have low autocorrelation and the linear complexity satisfies the requirements of cryptography.
\vspace*{0.1cm}
~
\end{abstract}

\section{Introduction}
A sequence $\textbf{u}=\{u_i\}_{i=0}^{\infty}$, if $\textbf{u}$ satisfies $u_{i+N}=u_i$, where $u_i\in\{0,1\}$, is called a binary sequence of period $\emph{N}$. The set $U=\{0\leq i <N:u_i=1\}$ is called the\emph{characteristic set} of $\textbf{u}$. If $|\emph{U}|=N/2$ for even N or $|\emph{U}|=(N\pm 1)/2$ for odd $\emph{N}$, where $|\emph{U}|$ denotes the cardinality of $\emph{U}$, then such a sequence \textbf{u} is called a balanced sequence.
Let $\textbf{u}=\{u_i\}_{i=0}^{\infty}$ and $\textbf{v}=\{v_i\}_{i=0}^{\infty}$ be two sequences of period $N$. The periodic correlation between them is defined by
\begin{equation}\label{eq01}
R_{\textbf{u}, \textbf{v}}(\tau) = \sum_{i = 0}^{N-1}(-1)^{\textbf{u}(i)+\textbf{v}(i+\tau)}, 0\leq\tau<N,
\end{equation}
where the addition $t+\tau$ is performed modulo $N$. $R_{\textbf{u}, \textbf{v}}(\tau)$ is called the (periodic) cross-correlation function of $\textbf{u}$ and $\textbf{v}$. If $\textbf{u} = \textbf{v}$, $R_{\textbf{u}, \textbf{v}}(\tau)$ is called the (period) autocorrelation function of $\textbf{u}$, denoted by $R_\textbf{u}(\tau)$ for short \cite{Ref3}.

According to the remainder of $N$ modulo $4$, the optimal values of out-of-phase autocorrelations of binary sequences are classified into four types as follows:
(1) $R_\textbf{a}(\tau)=-1$ if $N\equiv3 \bmod 4$; (2) $R_\textbf{a}(\tau)\in\{-2, 2\}$ if $N\equiv2 \bmod 4$; (3) $R_\textbf{a}(\tau)\in\{1, -3\}$ if $N\equiv1 \bmod 4$; (4) $R_\textbf{a}(\tau)\in\{0, -4,4\}$ if $N\equiv0 \bmod 4$, where $0<\tau<N$. For the second case, it is exactly the autocorrelation of the sequence we constructed. Binary sequences with low correlation have very significant applications in communication systems, radar and cryptography \cite{Ref2, Ref4}. Sequences should have low autocorrelation to eliminate the effect of multipath, and low cross-correlation to extract the desired user's signal from the rest of users.

In cryptographic applications, the linear complexity of a sequence is considered as the most important property. Generally speaking, a sequence with large linear complexity is favorable for cryptography to resist the well-known Berlekamp-Massey algorithm \cite{Ref6, Ref7}, and the sequence can be recovered easily if the linear complexity is less than half the period \cite{Ref8}.  It is worth to know the linear complexity of a sequence before applying them in any applications.

Many sequences with these good properties have interleaved structure, interleaved technique is widely used to analyse and design sequences \cite{Ref6}. In $2008$, based on the interleaved structure, Yu and Gong \cite{Ref18} presented a family of binary sequences of period $4N$ with optimal autocorrelation magnitude, i.e., $R_{\emph{\textbf{s}}}(\tau)\in \{0,\pm4\}$ for all $0<\tau<n$, and they also determined the linear complexity of the proposed sequence. Later, Tang and Gong \cite{Ref15} generalized the sequences in \cite{Ref18} and obtained more binary sequences of period $4N$ with optiaml autocorrelation value.

In this paper, using the interleaved technique, We construct the binary sequence with the otimal autocorrelation of period $2N$, and we calculate its autocorrelation value and distribution, and give a lower bound of linear complexity. Results show that these sequences have low autocorrelation and the linear complexity satisfies the requirements of cryptography.

This paper is organized as follows. Section $2$ introduces some related definitions and lemmas which would be used later. In Section $3$, We first give the interleaved structure, then calculate the autocorrelation value and its distribution of the sequences. In Section $4$, we give a lower bound of linear complexity. In Section $5$, we give a Remark.  Conclusions are given in Section $6$.

\section{Preliminaries}
\subsection{Interleaved Sequence}
\begin{definition}\cite{Ref8}
 Let $\{a_0,a_1,\cdots,a_{T-1}\}$ be a set of $T$ sequences of period $N$. An $N \times T$ matrix $U$ is formed by placing the sequence $a_i$ on the $i$th column, where $0 \leq i \leq T-1$. Then one can obtain an interleaved sequence $\textbf{u}$ of period $NT$ by concatenating the successive rows of the matrix $U$. For simplicity, the interleaved sequence $\textbf{u}$ can be written as
$$\textbf{u}=\mathbf{I}(a_0,a_1,\cdots,a_{T-1}),$$
where $\mathbf{I}$ denotes the interleaved operator.
\end{definition}

Let $\textbf{\emph{s}} = (\textbf{\emph{s}}(i))_{i = 0}^\infty$ be a sequence over a field $\mathbb{F}_2$. A polynomial of the form
$$f(x)=1+c_1x+c_2x^2+ \cdots +c_rx^r\in \mathbb{F}[x]$$
is called the characteristic polynomial of the sequence $\textbf{\emph{s}}$ if
$$\textbf{\emph{s}}(i)=c_1\textbf{\emph{s}}(i-1)+c_2\textbf{\emph{s}}(i-2)+ \cdots +c_r\textbf{\emph{s}}(i-r), \forall i \geq r.$$

Among all the characteristic polynomials of $\textbf{\emph{s}}$, the monic polynomial $m_\textbf{\emph{s}}(x)$ with the lowest degree is called its minimal polynomial. The linear complexity of $\textbf{\emph{s}}$ is defined as the degree of $m_\textbf{\emph{s}}(x)$, which is described as LC($\textbf{\emph{s}}$).

Let $\textbf{\emph{s}} = (\textbf{\emph{s}}(0),\textbf{\emph{s}}(1), \cdots , \textbf{\emph{s}}(n-1))$ be a binary sequence of period $n$ and define the sequence polynomial
\begin{equation}\label{eq02}
S(x) = \textbf{\emph{s}}(0)+\textbf{\emph{s}}(1)x+ \cdots +\textbf{\emph{s}}(n-1)x^{n-1}.
\end{equation}

Then, its minimal polynomial and linear complexity can be determined by Lemma $1$.

\begin{lemma}\cite{Ref15} Assume a sequence \textbf{\emph{s}} of period $n$ with sequence polynomial S(x) is defined by Equation~$\mathrm{(\ref{eq02})}$. Then
\begin{enumerate}
  \item  the minimal polynomial is $m_\textbf{\emph{s}}(x)= \frac{x^n-1}{\mathrm{gcd}(x^n-1,\textbf{\emph{s}}(x))};$
  \item the linear complexity is $\mathrm{LC}(\textbf{\emph{s}})=n-\mathrm{deg}(\mathrm{gcd}(x^n-1,S(x))),$
\end{enumerate}
where $\mathrm{gcd}(x^n-1,S(x))$ denotes the greatest common divisor of $x^n-1$ and $S(x)$.
\end{lemma}

\begin{lemma}\cite{Ref22} Let \textbf{a} be a binary sequence of period $n$, and $\textbf{\emph{s}}_\textbf{a}(x)$ be its sequence polynomial. Then
\begin{enumerate}
  \item $S_\textbf{b}(x) = x^{n-\tau}S_\textbf{a}(x), \; \textrm{if} \; \textbf{b} = L^\tau(\textbf{a})$;
  \item $S_\textbf{b}(x) = S_\textbf{a}(x)+\displaystyle\frac{x^n-1}{x-1}, \; \textrm{if} \; \textbf{b} \; is \; the \; complement \;\\ sequence \; of \; \textbf{a}$;
  \item $S_\textbf{u}(x) = S_\textbf{a}(x^2)+xS_\textbf{b}(x^2), \; \textrm{if} \; \textbf{u} = \mathbf{I}(\textbf{a},\textbf{b})$.
\end{enumerate}
\end{lemma}

\begin{lemma}\cite{Ref17} Let m be an integer. Correlation of sequences satisfies the following properties:
\begin{enumerate}
  \item $R_{L^m(\textbf a)\textbf b}(\tau)=R_{\textbf {ab}}(\tau-m)$;
  \item $R_{\textbf aL^m(\textbf b)(\tau)}=R_{\textbf {ab}}(\tau+m)$;
  \item $R_{\textbf{ab}}(\tau)=R_{\textbf{ab}}(\tau+N)=R_{\textbf{ba}}(N-\tau)$;
  \item $R_{\textbf{ab}}(\tau)+R_{\textbf{a} \bar{\textbf{b}}}(\tau)=R_{\textbf{ab}}(\tau)+R_{\bar{\textbf{a}}\textbf{b}}(\tau)=0$.
  \end{enumerate}
  \end{lemma}

\begin{lemma}The autocorrelation of $\textbf{u}=\mathbf{I}(\textbf{a},\textbf{b})$
\begin {equation}
R_\textbf{u}(\tau)=
\begin{cases}
R_\textbf{a}(\tau/2)+R_\textbf{b}(\tau/2) & \mbox{if }\tau \mbox{ is even}, \\
R_{\textbf{ab}}(\frac{\tau-1}{2})+R_{\textbf{ba}}(\frac{\tau+1}{2}) & \mbox{if }\tau \mbox{ is odd}.
\end{cases}
\end{equation}
\end{lemma}
\noindent\textbf{Proof}
For the case $\tau$ is even, we can know the location of sequence $\textbf{a}$ is replaced by $L^{\frac{\tau}{2}}(\textbf{a})$, and the location of sequence $\textbf{b}$ is replaced by $L^{\frac{\tau}{2}}(\textbf{b})$, in other words, $\mathbf{I}(\textbf{a},\textbf{b})$ becomes $\mathbf{I}(L^{\frac{\tau}{2}}(\textbf{a}),L^{\frac{\tau}{2}}(\textbf{b}))$, so by the definition, we have $R_\textbf{u}(\tau)=R_\textbf{a}(\frac{\tau}{2})+R_\textbf{b}(\frac{\tau}{2})$.

For the case $\tau$ is odd, we can know the location of sequence $\textbf{a}$ is replaced by $L^{\frac{\tau-1}{2}}(\textbf{b})$, and the location of sequence $\textbf{b}$ is replaced by $L^{\frac{\tau+1}{2}}(\textbf{a})$, in other words, $\mathbf{I}(\textbf{a},\textbf{b})$ becomes $\mathbf{I}(L^{\frac{\tau-1}{2}}(\textbf{b}),L^{\frac{\tau+1}{2}}(\textbf{a}))$, so by the definition we have $R_\textbf{u}(\tau)=R_{\textbf{ab}}(\frac{\tau-1}{2})+R_{\textbf{ba}}(\frac{\tau+1}{2})$. Hence, we have completed the proof of Lemma $4$.

Let $N$ be a prime and $\beta$ a primitive element of the integer residue ring $Z_N=\{0,1,\cdots,N-1\}$, such that for any $j\in Z_N \backslash \{0\}$, there exists an integer $k$ satisfying $j=\beta^k$. Denote by $D_0$ a multiplicative subgroup generated by $\beta^4$, i.e., $D_0=\{\beta^{4k}:0\leq k<f\}$. Then, $Z_N^*=Z_N \backslash \{0\}$ can be decomposed as $Z_N^*=\cup_{j=0}^3{D_j}$, where $D_j=\{\beta^{4k+j}:0\leq k<f\}$ is called the cyclotomic class $j$ of order $4$.

\begin{lemma}\cite{Ref3} Let $\textbf{u}$ and $\textbf{v}$ be two binary sequences with characteristic sets $D_0\cup D_1$ and $D_1\cup D_2$ respectively, and $\textbf{u}(0)=0$ and $\textbf{v}(0)=0$. Then we have
\begin{eqnarray*}
 R_\textbf{u}(\tau)=\left\{ \begin{array}{lll}
   N& \tau=0,\\
   -1-2y& \tau \in D_0\cup D_2,\\
   -1+2y& \tau \in D_1\cup D_3,
   \end{array} \right.
   \end{eqnarray*}
\begin{eqnarray*}
   R_{\textbf{v}}(\tau)= \left\{ \begin{array}{lll}
   N& \tau=0,\\
   -1+2y& \tau \in D_0\cup D_2,\\
   -1-2y& \tau \in D_1\cup D_3,
   \end{array} \right.
   \end{eqnarray*}
and
\begin{eqnarray*}
R_{\textbf{u} \textbf{v}}(\tau)=\left\{ \begin{array}{ll}
  -3& \tau \in D_2,\\
  1& otherwise.
  \end{array} \right.
   \end{eqnarray*}
   \end{lemma}

\begin{lemma}\cite{Ref21} Let $S_{\textbf{u}}(x)$ and $S_{\textbf{v}}(x)$ be the sequence polynomials of sequences $\textbf{u}$ and $\textbf{v}$, $\alpha$ a primitive $N$th root of unity over the field $GF(2^m)$, that is the splitting field of $x^N-1$. Then
\begin{eqnarray*}
S_{\textbf{u}}(\alpha^j)= \left\{ \begin{array}{lll}
S_{\textbf{u}}(\alpha)& j\in D_0,\\
S_{\textbf{v}}(\alpha)& j\in D_1,\\
1+S_{\textbf{u}}(\alpha)& j\in D_2,\\
1+S_{\textbf{v}}(\alpha)& j\in D_3,\\
\frac{N-1}{2}\pmod 2& j=0,
\end{array} \right.
   \end{eqnarray*}
\begin{eqnarray*}
S_{\textbf{v}}(\alpha^j)= \left\{ \begin{array}{lll}
S_{\textbf{v}}(\alpha)& j\in D_0,\\
1+S_{\textbf{u}}(\alpha)& j\in D_1,\\
1+S_{\textbf{v}}(\alpha)& j\in D_2,\\
S_{\textbf{u}}(\alpha)& j\in D_3,\\
\frac{N-1}{2}\pmod 2& j=0,
\end{array} \right.
\end{eqnarray*}
\end{lemma}
where
\begin{center}
$S_{\textbf{u}}(x)=\sum_{i\in D_0\cup D_1} x^{i},$\\
$S_{\textbf{v}}(x)=\sum_{i\in D_1\cup D_2} x^{i}.$
\end{center}

\begin{lemma}\cite{Ref21} $S_{\textbf{u}}(\alpha)\in \{0,1\}$ if and only if $2\in D_0$.
\end{lemma}

\begin{lemma} $S_{\textbf{v}}(\alpha)\in \{0,1\}$ if and only if $2\in D_0$.
\end{lemma}
\noindent\textbf{Proof}
Since $(D_0,\cdot)$ is a group, we have $qD_0=D_0$ and $q^{-1}\in D_0$ for any $q\in D_0$. Hence
\begin{eqnarray*}
& &S_{\textbf{v}}(\alpha^q)\\
&=&\sum_{i\in D_1\cup D_2}{\alpha^{qi}}\\
&=&\sum_{y\in D_1\cup D_2}{\alpha^y}\\
&=&S_{\textbf{v}}(\alpha).
\end{eqnarray*}
Since the characteristic of the field $GF(2^m)$ is $2$, it follows that $(S_{\textbf{v}}(\alpha))^2=S_{\textbf{v}}({\alpha}^2)$. From the above, we have $S_{\textbf{v}}(\alpha^2)=S_{\textbf{v}}(\alpha)$ if and only if $2\in D_0$, that is to say $S_{\textbf{v}}(\alpha)\in \{0,1\}$ if and only if $2\in D_0$. So we have completed the proof of Lemma $8$.

\section{New Construction Method and the Autocorrelation Values and Distribution }
In this section, assume that $\textbf{u}$ and $\textbf{v}$ are two binary sequences with characteristic sets $D_i\cup D_j$ and $D_j\cup D_l$ respectively, and we define
\begin{align*}
\textbf{u}'(t)=\left\{ \begin{array}{lll}
 u(t)& t\neq 0,\\
 1& t=0.
 \end{array} \right.
 \end{align*}
 \begin{align*}
\textbf{v}'(t)=\left\{ \begin{array}{lll}
 v(t)& t\neq 0,\\
 1& t=0.
 \end{array} \right.
 \end{align*}
 we proposes one new way to construct sequences, then we calculate its autocorrelation values and distribution.

\subsection{New Construction Method and Sequence Correspondence}
Let $N=4f+1$, where $f=y^2$ is odd and $N\geq 5$, and sequences $\textbf{u}$ and $\textbf{v}$ are the same as before, new construction as the following:
\begin{equation}
\textbf{\emph{s}}=\mathbf{I}(\textbf{u},L^{(N+1)/2}\textbf{v}),
\end{equation}
\begin{equation}
\textbf{\emph{s}}'=\mathbf{I}(\textbf{u}',L^{(N+1)/2}\textbf{v}).
\end{equation}

Compare the above two constructions, we can know that when $\textbf{\emph{s}}'=\mathbf{I}(\textbf{u}',L^{(N+1)/2}\textbf{v})$, the sequence is a balanced sequence, then we give the autocorrelation value for different characteristic sets for $\textbf{\emph{s}}'$, which is similarly demonstrated when $\textbf{\emph{s}}=\mathbf{I}(\textbf{u},L^{(N+1)/2}\textbf{v})$. Below we calculate their autocorrelation of $\textbf{\emph{s}}'$ in different situations.

\subsection{the Autocorrelation Values and Distribution}
$\mathbf{Case \; 1.}$ $(i,j,l)=(0,1,2), and (i,j,l)=(2,1,0)$.

\noindent\textbf{Proof}
When $(i,j,l)=(0,1,2)$, we know that the characteristic sets of $\textbf{u}'$ and $\textbf{v}$ are $D_0\cup D_1$ and $D_1\cup D_2$ respectively.
Let $\beta$ be a primitive element of the integer residue ring $Z_N=\{0,1,\cdots,N-1\}$, such that for any $j\in Z_N \backslash \{0\}$, there exists an integer $k$ satisfying $j=\beta^k$. We have $1=\beta^{4f}$, since $ord(\beta)=N-1$ and $f$ is odd, $f$ can be empressed by $f=2t+1$, $t$ is an integer, so $(\beta)^{4t+2}=-1$, we have $-1\in D_2$. For the sequences $\textbf{u}$ and $\textbf{v}$, since $-1\in D_2$, by Lemmas $3(3)$ and $5$, we know:
\begin{align}
 R_{\textbf{v} \textbf{u}}(\tau)=\left\{ \begin{array}{lll}
 -3& \tau \in D_0,\\
 1& otherwise.
 \end{array} \right.
 \end{align}

For the autocorrelation values of sequences $\textbf{u}'$ and $\textbf{v}'$, since the only difference between sequences $\textbf{u}$ and $\textbf{u}'$ is $\textbf{u}'(0)=1$, we only need to know the values of $\textbf{u}'(\tau)$ and $\textbf{u}'(N-\tau)$, since $-1\in{D_2}$, $\textbf{u}'(\tau)$ and $\textbf{u}'(N-\tau)$ take different values of $0$ and $1$, we have $(-1)^{\textbf{u}'(0)+\textbf{u}'(\tau)}+(-1)^{\textbf{u}'(0)+\textbf{u}'(N-\tau)}=0$, so we have $R_{\textbf{u}}(\tau)=R_{\textbf{u}'}(\tau)$, similarly we know $R_{\textbf{v}}(\tau)=R_{\textbf{v}'}(\tau)$, for the cross-correlation value of sequences $\textbf{u}$ and $\textbf{v}'$, we only to know $\textbf{v}(\tau)$ and $\textbf{v}(N-\tau)$, then by Lemma $5$ and Equation $(6)$, we have:
\begin{align*}
R_{\textbf{u}'\textbf{v}}(\tau)=\left\{ \begin{array}{ll}
  3& \tau \in D_1,\\
  -1& otherwise.
  \end{array} \right.
    \end{align*}
\begin{align*}
R_{\textbf{v}\textbf{u}'}(\tau)=\left\{ \begin{array}{ll}
  3& \tau \in D_3,\\
  -1& otherwise.
  \end{array} \right.
   \end{align*}

By Lemma $4$, we know:
\begin {equation}
R_\textbf{\emph{s}}(\tau)=
\begin{cases}
R_\textbf{u}(\tau/2)+R_{\textbf{v}'}(\tau/2)& \tau \mbox{ is even},\\
R_{\textbf{u}'\textbf{v}}(\frac{\tau+N}{2})+R_{\textbf{v}\textbf{u}'}(\frac{\tau+N}{2})& \tau \mbox{ is odd}.
\end{cases}
\end{equation}

Then by Lemma $5$, we have
\begin{eqnarray}
 R_\textbf{\emph{s}}(\tau)=\left\{ \begin{array}{lll}
 2N& \tau=0,\\
 -2& \tau \mbox{ is even},\tau \neq 0,\\
 -2& \tau \mbox{ is odd}, \frac{\tau+N}{2} \in D_0,\\
 2& \tau \mbox{ is odd}, \frac{\tau+N}{2} \in D_1 ,\\
 -2& \tau \mbox{ is odd}, \frac{\tau+N}{2} \in D_2,\\
 2& \tau \mbox{ is odd}, \frac{\tau+N}{2} \in D_3,\\
 -2& \tau=N.\\
 \end{array} \right.
 \end{eqnarray}

When $(i,j,l)=(2,1,0)$, compared to $(i,j,l)=(0,1,2)$, their difference is that the position of the base sequence has changed, so the autocorrelation values and cross-correlation values of the base sequences are unchanged, then by Equation $(7)$, the autocorrelation values of the sequence $\textbf{\emph{s}}$ is the same as above.\\
$\mathbf{Case \; 2.}$ $(i,j,l)=(1,2,3),(i,j,l)=(3,2,1)$.

\noindent\textbf{Proof}
According to the proof of case $1$, we can know that when the characteristic sets (i,j,l)=(1,2,3) and (i,j,l)=(3,2,1), the autocorrelation values of the sequences are equal, and we can know the autocorrelation values of the base sequence:
\begin{eqnarray*}
 R_{\textbf{u}'}(\tau)=\left\{ \begin{array}{lll}
   N& \tau=0,\\
   -1+2y& \tau \in D_0\cup D_2,\\
   -1-2y& \tau \in D_1\cup D_3,
   \end{array} \right.
   \end{eqnarray*}
\begin{eqnarray*}
   R_{\textbf{v}}(\tau)= \left\{ \begin{array}{lll}
   N& \tau=0,\\
   -1-2y& \tau \in D_0\cup D_2,\\
   -1+2y& \tau \in D_1\cup D_3,
   \end{array} \right.
   \end{eqnarray*}

The cross-correlation values of the base sequence:
\begin{align*}
R_{\textbf{u}'\textbf{v}}(\tau)=\left\{ \begin{array}{ll}
  3& \tau \in D_2,\\
  -1& otherwise.
  \end{array} \right.
    \end{align*}
\begin{align*}
R_{\textbf{v}\textbf{u}'}(\tau)=\left\{ \begin{array}{ll}
  3& \tau \in D_0,\\
  -1& otherwise.
  \end{array} \right.
   \end{align*}

Then by Equation $(7)$, we have
\begin{eqnarray}
 R_\textbf{\emph{s}}(\tau)=\left\{ \begin{array}{lll}
 2N& \tau=0,\\
 -2& \tau \mbox{ is even},\tau \neq 0,\\
 2& \tau \mbox{ is odd}, \frac{\tau+N}{2} \in D_0,\\
 -2& \tau \mbox{ is odd}, \frac{\tau+N}{2} \in D_1 ,\\
 2& \tau \mbox{ is odd}, \frac{\tau+N}{2} \in D_2,\\
 -2& \tau \mbox{ is odd}, \frac{\tau+N}{2} \in D_3,\\
 -2& \tau=N.\\
 \end{array} \right.
 \end{eqnarray}

The above is the autocorrelation function of sequences $\textbf{\emph{s}}'$, the sequences $\textbf{\emph{s}}$ can be similarly demonstrated, in the case of the same characteristic set, except for $\tau=N$, the autocorrelation values are opposite to each other, in other cases, the autocorrelation values are the same, and its proof is omitted for simplicity. Next we give the autocorrelation distribution of the sequence $\textbf{\emph{s}}$.

\begin{theorem}The autocorrelation value distribution of $\textbf{\emph{s}}=\mathbf{I}(\textbf{u}',L^{(N+1)/2}\textbf{v})$:
\begin{eqnarray}
 R_\textbf{\emph{s}}(\tau)=\left\{ \begin{array}{lll}
 2N&  1\,\,time,\\
   -2&  \frac{7N-3}{4}\,\,time,\\
   2&  \frac{N-1}{4}\,\,times.
 \end{array} \right.
 \end{eqnarray}
When $\textbf{\emph{s}}=\mathbf{I}(\textbf{u},L^{(N+1)/2}\textbf{v})$:
\begin{eqnarray}
 R_\textbf{\emph{s}}(\tau)=\left\{ \begin{array}{lll}
 2N&  1\,\,time,\\
   -2&  \frac{7N-7}{4}\,\,time,\\
   2&  \frac{N+3}{4}\,\,times.
 \end{array} \right.
 \end{eqnarray}
 \end{theorem}

\noindent\textbf{Proof}
Let $Q$ denote the quadratic residual set of modulo $N$, $P$ be a quadratic non-residual set for modulo $N$, then $Q=D_0\cup D_2$, $P=D_1\cup D_3$, since $-1\in D_2$, and $N$ is odd, if $a\in Q$ is odd(even), then $N-a\in Q$, and $N-a$ is even(odd). Thus half elements in $Q$ are even(odd). So is $P$. Then by Equations $(8)$ and $(9)$, the Equation $(10)$ is proved. When $\textbf{\emph{s}}=\mathbf{I}(\textbf{u},L^{(N+1)/2}\textbf{v})$ similarly demonstrated. So Theorem $1$ is proved.

\section{A Lower Bound of Linear Complexity}
Since $N=4f+1$, $2\notin D_0 \cup D_2$, then we only consider the case $2\notin D_0$, $\textbf{u}'$ and $\textbf{v}$ are two binary sequences with characteristic sets $D_0\cup D_1$ and $D_1\cup D_2$ respectively, then we have:
\begin{corollary}The minimal polynominls of $\textbf{u}'$ and $\textbf{v}$:
\begin{enumerate}
  \item $m_{\textbf{u}'}(x)=x^N-1$;
  \item $m_{\textbf{v}}(x)=\displaystyle\frac{x^N-1}{x-1}$.
  \end{enumerate}
  \end{corollary}
\noindent\textbf{Proof}
Let $\alpha$ be a primitive $N$th root over $GF(2^m)$ as before. Since $N=4f+1$, $2\notin D_0$, by Lemmas $7$ and $9$, $S_{\textbf{u}'}(\alpha) \notin \{0,1\}$ and $S_{\textbf{v}}(\alpha) \notin \{0,1\}$, so $\{\alpha^j: 0<j\leq {N-1}\}$ are not the roots of $S_{\textbf{u}'}(\alpha)$ and $S_{\textbf{v}}(\alpha)$. Since $S_{\textbf{u}'}(\alpha^0)=(\frac{N-1}{2}+1)=1 \pmod 2$, $S_{\textbf{v}}(\alpha^0)=\frac{N-1}{2}=0 \pmod 2$, we have $\textrm{gcd}(S_{\textbf{u}'}(x),x^{N}-1)=1$, $\textrm{gcd}(S_{\textbf{v}}(x),x^{N}-1)=x-1$, then by the Lemma $1$, we have completed the proof.

\begin{corollary}When $2\in D_1$, we have $S_{\textbf{v}}(x^2)=S_{\textbf{u}'}(x)$, and $2\in D_3$, we have $S_{\textbf{u}'}(x^2)=S_{\textbf{v}}(x)$.
  \end{corollary}
\noindent\textbf{Proof}
When $2\in D_1$,
\begin{eqnarray*}
& &S_{\textbf{v}}(x^2)\\
&=&\sum_{i\in D_1\cup D_2} x^{2i}\\
&=&\sum_{g\in D_2\cup D_3} x^{g}\\
&=&S_{\textbf{u}}(x);
\end{eqnarray*}
 When $2\in D_3$,
\begin{eqnarray*}
& &S_{\textbf{u}}(x^2)\\
&=&1+\sum_{i\in D_0\cup D_1} x^{2i}\\
&=&1+\sum_{h\in D_0\cup D_3} x^{h}\\
&=&S_{\textbf{v}}(x).
\end{eqnarray*}
We have completed the proof.

\begin{theorem}
Let $\textbf{\emph{s}}'$ be the interleaved sequence of period $2N$ as before and $S_{\textbf{\emph{s}}'}(x)$ is the sequence polynomial of $\textbf{\emph{s}}'$, $\textbf{\emph{s}}'=\mathbf{I}(\textbf{u}',L^{(N+1)/2}\textbf{v})$. Then the linear complexity $LC(\textbf{\emph{s}}')$ is bounded by $LC(\textbf{\emph{s}}') \geq 2N-4$, when $\textbf{\emph{s}}=\mathbf{I}(\textbf{u},L^{(N+1)/2}\textbf{v})$, the linear complexity $LC(\textbf{\emph{s}})$ is bounded by $LC(\textbf{\emph{s}}) \geq 2N-5$.
\end{theorem}
\noindent\textbf{Proof}
Let $S_{\textbf{u}'}(x)$ and $S_{\textbf{v}}(x)$ be the sequence polynomials of $\textbf{u}'$ and $\textbf{v}$, the characteristic sets $(i,j,l)=(0,1,2),(1,2,3),(2,1,0)$ or $(3,2,1)$, $\alpha$ be a primitive $N$th root of unity over the field $GF(2^m)$. Let us take the characteristic set (i,j,l)=(0,1,2) as an example, and other cases can be similarly proved. By Lemma $2(3)$, we have
\begin{eqnarray*}
 & &\textrm{gcd}(S_{\textbf{\emph{s}}'}(x),x^{2N}-1)\\
 &=&\textrm{gcd}((S_{\textbf{u}'}(x^2)+x^{\frac{N+1}{2}}S_{\textbf{v}}(x^2),x^{2N}-1).
\end{eqnarray*}

Since $S_{\textbf{\emph{s}}'}(\alpha^0)=1$, then $\textrm{gcd}(S_{\textbf{\emph{s}}'}(\alpha^0),((\alpha^0)^{2N}-1))=1$, so $x-{\alpha^0}$ is not a common factor. Then we consider whether $\{\alpha^j:j\in Z_N^*\}$ is the common root of $S_{\textbf{\emph{s}}'}(x)$ and $x^{2N}-1)$.

When $2\in D_1$, by Corollary $3$, we have
 \begin{eqnarray*}
 & &\textrm{gcd}(S_{\textbf{u}'}(x^2)+x^{\frac{N+1}{2}}S_{\textbf{v}}(x^2), x^{2N}-1)\\
 &=&\textrm{gcd}(S_{\textbf{u}'}(x)(S_{\textbf{u}'}(x)+x^{\frac{N+1}{2}}),x^{2N}-1).
\end{eqnarray*}

Since $\textrm{gcd}(S_{\textbf{u}'}(x),x^{N}-1)=1$, so we only consider $\textrm{gcd}(S_{\textbf{u}'}(x)+x^{\frac{N+1}{2}},x^{2N}-1)$, for $\{\alpha^j:j\in D_0\}$, $(\alpha)^{\frac{j.(N+1)}{2}}$ are not equal to each other, then by Lemma $6$,for $\{\alpha^j:j\in D_0\}$, $S_{\textbf{u}'}(x)$ is the same value, we can know in the set $\{\alpha^j:j\in D_0\}$, only one number may be the root of the equation $S_{\textbf{u}'}(x)+x^{\frac{N+1}{2}}=0$. Similarly, there may be only three roots for the other three cases, so the linear complexity is bounded by $LC(\textbf{\emph{s}}') \geq 2N-4$.

When $2\in D_3$, then by Corollary $3$, we have
\begin{eqnarray*}
 & &\textrm{gcd}(S_{\textbf{u}'}(x^2)+x^{\frac{N+1}{2}}S_{\textbf{v}}(x^2), x^{2N}-1)\\
 &=&\textrm{gcd}(S_{\textbf{v}}(x)(1+x^{\frac{N+1}{2}}S_{\textbf{v}}(x)),x^{2N}-1).
\end{eqnarray*}

Since $\textrm{gcd}(S_{\textbf{v}}(x),x^{N}-1)=1$, we only consider $\textrm{gcd}(1+x^{\frac{N+1}{2}}S_{\textbf{v}}(x),x^{2N}-1)$, according to the above proof, for $\{\alpha^j:j\in D_0\}$, $S_{\textbf{v}}(x)$ is the same value, we can know in the set $\{\alpha^j:j\in D_0\}$, only one number may be the root of the equation $1+x^{{\frac{N+1}{2}}}S_{\textbf{v}}(x)$. Similarly, there may be only three roots for the other three cases, so the linear complexity is bounded by $LC(\textbf{\emph{s}}') \geq 2N-4$. Hence, we have completed the proof of Theorem $2$.

When $\textbf{\emph{s}}=\mathbf{I}(\textbf{u},L^{(N+1)/2}\textbf{v})$, the proof of linear complexity is the same as for $\textbf{\emph{s}}'=\mathbf{I}(\textbf{u}',L^{(N+1)/2}\textbf{v})$, the only difference between $\textbf{\emph{s}}=\mathbf{I}(\textbf{u},L^{(N+1)/2}\textbf{v})$ and $\textbf{\emph{s}}'=\mathbf{I}(\textbf{u}',L^{(N+1)/2}\textbf{v})$ is that $1$ is the root of $\textbf{\emph{s}}=\mathbf{I}(\textbf{u},L^{(N+1)/2}\textbf{v})$, so the linear complexity is bounded by $LC(\textbf{\emph{s}}) \geq 2N-5$.

\section{Remark}
$\textbf{\emph{s}}$ and $\textbf{\emph{s}}'$ possess optiaml autocorrelation if and only if its characteristic set is an almost difference set, sequences $\textbf{\emph{s}}$ and $\textbf{\emph{s}}'$ can also be obtained by the results in \cite{Ref1}.

\section{Conclusion}
In this paper, we use the interleaving technique to construct a binary sequence with the optimal autocorrelation of period $2N$. From the section $3$, we can conclude that this sequence is optimal, the sequence has low autocorrelation values, and we give the distribution of the autocorrelation values. Especially, in section $4$, based on the discussion of roots of the sequence polynomials over the field $\textrm{GF}(2^m)$, we give a lower bound of linear complexity, the linear complexity can be reached $2N-4$, far larger than half of a period. Results show that these sequences have low autocorrelation and the linear complexity satisfies the requirements of cryptography.

\section*{Acknowledgments}
The project is supported by the open fund of Fujian Provincial Key Laboratory of Network Security and Cryptology Research Fund (Fujian Normal University)
(No.15002), the Natural Science Fund of Shandong Province (No.ZR2014FQ005), the National Natural Science Foundations of China(No.61170319)and the Fundamental Research Funds for the Central Universities (No.11CX04056A, 15CX08011A,15CX05060A).

\addcontentsline{toc}{chapter}{\protect\numberline{}{REFERENCES}}

\noindent {\bf Shidong Zhang} biography.
 Shidong Zhang was born in 1992 in Shandong Province of China. He was graduated from Jining University . He will study for a postgraduate degree at China University of Petroleum in 2016. And his tutor is Tongjiang Yan. Email:zhangshdo1992@163.com\\

\noindent {\bf Tongjiang Yan} biography.
Tongjiang Yan was born in 1973 in Shandong Province of China. He was graduated from the Department of Mathematics, Huaibei Coal-industry Teachers College, China, in 1996. In 1999, he received the M.S. degree in mathematics from the Northeast Normal University, Lanzhou, China. In 2007, he received the Ph.D. degree in Xidian University. He is now a professor of China University of Petroleum. His research interests include cryptography and algebra. Email:yantoji@163.com \\
\end{document}